\documentclass{elsart}

\begin{document}

\begin{frontmatter}
  \title{Molecular dynamics study of a classical two-dimensional
    electron system:\\ Positional and orientational orders}
  \author{Satoru Muto\thanksref{e-mail}},
  \author{Hideo Aoki}
  \address{Department of Physics, University of Tokyo, Hongo,
    Tokyo 113-0033, Japan}
  \thanks[e-mail]{Corresponding author.
    E-mail: mutou@cms.phys.s.u-tokyo.ac.jp}
  \begin{abstract}
    Molecular dynamics simulation is used to investigate the
    crystallization of a classical two-dimensional electron system, in
    which electrons interact with the Coulomb repulsion.  From the
    positional and the orientational correlation functions, we have
    found an indication that the solid phase has a quasi-long-range
    (power-law correlated) positional order and a long-range
    orientational order.  This implies that the long-range $1/r$
    system shares the absence of the true long-range crystalline order
    at finite temperatures with short-range ones to which Mermin's
    theorem applies.  We also discuss the existence of the ``hexatic''
    phase predicted by the Kosterlitz-Thouless-Halperin-Nelson-Young
    theory.
  \end{abstract}
\end{frontmatter}

\section{Introduction}
More than 60 years ago, Wigner pointed out that an electron system
will crystallize due to the Coulomb repulsion for low enough densities
(Wigner crystallization) \cite{wigner}.  Although quantum effects play
an essential role in a degenerate electron system, the concept of
Wigner crystallization can be generalized to a classical case where
the Fermi energy is much smaller than the thermal energy.  A classical
two-dimensional (2D) electron system is wholly specified by the
dimensionless coupling constant $\Gamma$, the ratio of the Coulomb
energy to the kinetic energy.  Here $\Gamma \equiv (e^{2}/4\pi\epsilon
a)/k_{B}T$, where $e$ is the charge of an electron, $\epsilon$ the
dielectric constant of the substrate, $a$ the mean distance between
electrons and $T$ the temperature.  For $\Gamma \ll 1$ the system will
behave as a gas while for $\Gamma \gg 1$ as a solid.  Experimentally,
Grimes and Adams \cite{grimes-adams} succeeded in observing a
transition from a liquid to a triangular lattice in a classical 2D
electron system on a liquid-helium surface around $\Gamma_{c} = 137
\pm 15$, which is in good agreement with numerical simulations
\cite{gann,morf,kalia,bedanov,schweigert}.

On the theoretical side, two conspicuous points have been known for 2D
systems: (i) Mermin's theorem dictates that no true long-range
crystalline order is possible at finite $T$ in the thermodynamic limit
\cite{mermin}.  To be precise, the $1/r$ Coulomb interaction is too
long ranged to apply Mermin's arguments directly.  Although there have
been some theoretical attempts \cite{chakravarty,alastuey} to extend
the theorem to the Coulomb case, no rigorous proof has been attained.
(ii) A theory due to Kosterlitz, Thouless, Halperin, Nelson, and Young
(KTHNY) predicts that a ``hexatic'' phase, characterized by a
short-range positional order and a quasi-long-range orientational
order, exists between a liquid phase and a solid phase
\cite{halperin-nelson}.  Because the KTHNY theory is based on various
assumptions and approximations, its validity should be tested by
numerical methods such as a molecular dynamics (MD) simulation.
Several authors have applied numerical methods to classical 2D electron
systems, but they arrived at different conclusions on the KTHNY
prediction \cite{morf,kalia,bedanov,schweigert}.

In order to address both of the above problems, the most direct way is
to calculate the positional and the orientational correlation
functions, which is exactly the motivation of the present study
\cite{footnote1}.

\section{Numerical Method}
A detailed description of the simulation is given elsewhere
\cite{muto}, so we only recapitulate it. We consider a rectangular
area with a rigid uniform neutralizing positive background in
periodic boundary conditions.  The aspect ratio of the rectangle is
taken to be $L_{y}/L_{x}=2/\sqrt{3}$, which can accommodate a perfect
triangular lattice \cite{bonsall}.  The Ewald summation method is used
to take care of the long-range nature of the $1/r$ interaction.  We
have employed Nos\'{e}-Hoover's canonical MD method \cite{nose,hoover}
to incorporate temperature accurately.

The system is cooled or heated across the transition with a simulated
annealing method.  The results presented here are for $N=900$
electrons with MD runs with 30,000--110,000 time-steps for each value
of $\Gamma$.  The correlation functions and other quantities are
calculated for the last $\sim$20,000 time-steps of each run.

Following Cha and Fertig \cite{cha-fertig}, we define the positional
and the orientational correlation functions from which we identify the
ordering in each phase.  The positional correlation function is
defined by
\begin{displaymath}
  C(r) \equiv \langle \rho^{*}_{\bf G}({\bf r})
  \rho_{\bf G}({\bf 0}) \rangle
  = \left\langle \frac{{\displaystyle \sum_{i,j}}
    \delta (r - |{\bf r}_{i} - {\bf r}_{j}|)
    \frac{1}{6}{\displaystyle \sum_{{\bf G}}}
    e^{{\rm i}{\bf G} \cdot ({\bf r}_{i} - {\bf r}_{j})}}
  {{\displaystyle \sum_{i,j}}
    \delta (r - |{\bf r}_{i} - {\bf r}_{j}|)} \right\rangle ,
\end{displaymath}
where ${\bf G}$ is a reciprocal vector of the triangular lattice with
the summation taken over the six ${\bf G}$'s that give the first peaks
of the structure factor [inset (a) of Fig.~\ref{corr-solid}].
The orientational correlation function is defined by
\begin{displaymath}
  C_{6}(r) \equiv \langle \psi^{*}_{6}({\bf r})
  \psi_{6}({\bf 0}) \rangle
  = \left\langle \frac{{\displaystyle \sum_{i,j}}
    \delta (r - |{\bf r}_{i} - {\bf r}_{j}|)
    \psi^{*}_{6} ({\bf r}_{i}) \psi_{6} ({\bf r}_{j})}
  {{\displaystyle \sum_{i,j}} \delta (r - |{\bf r}_{i} - {\bf r}_{j}|)}
  \right\rangle,
\end{displaymath}
where $\psi_{6} ({\bf r}) \equiv \frac{1}{n_{c}}\sum_{\alpha}^{\rm
n.n.}  e^{6{\rm i}\theta_{\alpha}({\bf r})}$ with
$\theta_{\alpha}({\bf r})$ being the angle of the vector connecting an
electron at ${\bf r}$ and the $\alpha$-th nearest neighbor with
respect to a fixed axis. The summation is taken over $n_{c}$ nearest
neighbors that are determined by the Voronoi construction
\cite{allen}.

\section{Results and Discussions}
Let us first look at the positional and the orientational correlation
functions in Fig.~\ref{corr-solid} for $\Gamma = 200$ and $\Gamma =
160$, for which the system is well in the solid phase.  The positional
correlation is seen to decay slowly, indicative of an algebraic
(power-law) decay at large distances.  The round-off in the
correlation function around half of the system size should be an
effect of the periodic boundary conditions.  The algebraic decay of
the positional correlation indicates that the 2D electron solid has
only a {\it quasi-long-range} positional order.  Thus we have obtained
a numerical indication that Mermin's theorem \cite{mermin} remains
applicable to the $1/r$ Coulomb interaction, which is consistent with
the analytical but approximate results obtained in
\cite{chakravarty,alastuey}.

On the other hand, the orientational correlation rapidly approaches a
constant, indicating a long-range orientational order.  Therefore,
while the 2D electron solid has no true long-range crystalline order,
we can say that it has a {\it topological} order.  From a snapshot of
the configuration [see inset (b) in Fig.~\ref{corr-solid}], we can see
that a long-range orientational order is preserved since defects (5-
or 7-fold disclinations, etc) tend to appear as dislocation (5-7
combination of disclinations) pairs, i.e., 5-7-5-7 disclination
quartets that only disturb the orientational correlation locally.
Here the coordination number is again determined from the Voronoi
construction.

Now we move on to the orientational correlation function (inset of
Fig.~\ref{discl-dist}) around the crystallization, which is obtained
by cooling the system from a liquid to a solid.  In between a
short-range orientational order for $\Gamma = 120$ and a long-range
one for $\Gamma = 140$, the correlation appears to decay algebraically
at $\Gamma = 130$ with an exponent approximately equal to unity, which
deviates from the upper bound of $1/4$ predicted by KTHNY \cite{footnote2}.
However, numerical difficulties arising from
finite-size and finite-time effects prevent us from drawing any
definite conclusion on the existence of the hexatic phase.  Namely, we
cannot rule out the possibility that the power-law behavior is an
artifact of insufficient equilibration.  In fact, the solid phase
persists down to $\Gamma = 130$ when the system is heated from a
solid, which is understandable if the solid-hexatic and the
hexatic-liquid transitions are of first and second order,
respectively.

The KTHNY theory is based on a picture that the hexatic-liquid
transition occurs through unbinding of disclination pairs.  To examine
if this is the case, we have calculated a defect-defect correlation
function (Fig.~\ref{discl-dist}), which we define as a distribution of
7-fold coordinated electrons with respect to a 5-fold coordinated
electron.  The correlation function exhibits no qualitative difference
between $\Gamma = 120$ and $\Gamma = 130$, for which the disclinations
are not tightly bound as compared with $\Gamma = 140$.  If we look at
a snapshot for $\Gamma = 130$ (Fig.~\ref{snapshot}), we see some
domain structure as far as the present numerical simulation with
finite-size and finite-time restrictions are concerned.
A finite-size scaling will be an interesting future problem
if the transition into the hexatic phase is of second order.

We have also looked at a dynamical property, i.e., the power spectral
density of the velocity (Fig.~\ref{velocity}), which is related to the
vibrational density of states and corresponds to the Fourier transform
of the velocity autocorrelation function via Wiener-Khinchin's
theorem.  While the difference between the solid and the liquid phases
appears in the spectrum around zero frequency, which is proportional
to the diffusion constant (finite in the liquid or vanishingly small
in the solid), we find a peak around the typical phonon frequency,
which, curiously enough, persists even in the liquid phase.  This
indicates that the liquid has well-defined local configurations
despite the short-range positional and orientational correlations.

\begin{figure}[p]
  \caption{The positional (the upper frame) and the orientational
    (the lower frame) correlation functions for $\Gamma = 200$ and
    $\Gamma = 160$.  The horizontal scale is the distance in units of
    the lattice constant of the triangular lattice. The structure
    factor for $\Gamma = 200$ is shown in inset (a), and a defect
    structure that is found for $\Gamma = 160$ is shown in inset (b),
    where 5-fold (7-fold) coordinated electrons are marked with open
    (solid) circles.}
  \label{corr-solid}
\end{figure}

\begin{figure}[p]
  \caption{Distribution of 7-fold coordinated electrons
    with respect to a 5-fold coordinated electron near the
    crystallization.
    The orientational correlation function is also shown in the inset.}
  \label{discl-dist}
\end{figure}

\begin{figure}[p]
  \caption{A snapshot of the electron configuration for $\Gamma = 130$.
    5-fold (7-fold) coordinated electrons are marked with open (solid)
    circles.}
  \label{snapshot}
\end{figure}
    
\begin{figure}[p]
  \caption{Power spectral density of the velocity for $\Gamma = 120$
    and $\Gamma = 140$.
    The plot for $\Gamma = 130$ is similar to that for $\Gamma = 120$.
    The frequency scale is for $n = 10^{12}/{\rm m^{2}}$.}
  \label{velocity}
\end{figure}


\begin{thebibliography}{99}
\bibitem{wigner}
  E.~Wigner, Phys.~Rev. {\bf 46}, 1002 (1934).
\bibitem{grimes-adams}
  C.~C.~Grimes and G.~Adams, Phys.~Rev.~Lett. {\bf 42}, 795 (1979).
\bibitem{bonsall}
  L.~Bonsall and A.~A.~Maradudin, Phys.~Rev. B {\bf 15}, 1959 (1977).
\bibitem{gann}
  R.~C.~Gann, S.~Chakravarty, and G.~V.~Chester, Phys.~Rev.
  B {\bf 20}, 326 (1979).
\bibitem{morf}
  R.~H.~Morf, Phys.~Rev.~Lett. {\bf 43}, 931 (1979).
\bibitem{kalia}
  R.~K.~Kalia, P.~Vashishta, and S.~W.~de Leeuw, Phys.~Rev.
  B {\bf 23}, 4794 (1981).
\bibitem{bedanov}
  V.~M.~Bedanov, G.~V.~Gadiyak, and Yu.~E.~Lozovik,
  Sov.~Phys. JETP {\bf 61}, 967 (1985).
\bibitem{schweigert}
  I.~V.~Schweigert, V.~A.~Schweigert, and F.~M.~Peeters,
  Phys.~Rev.~Lett. {\bf 82}, 5293 (1999).
\bibitem{mermin}
  N.~D.~Mermin, Phys.~Rev. {\bf 176}, 250 (1968).
\bibitem{chakravarty}
  S.~Chakravarty and C.~Dasgupta, Phys.~Rev. B {\bf 22}, 369 (1980).
\bibitem{alastuey}
  A.~Alastuey and B.~Jancovici, J.~Stat.~Phys. {\bf 24}, 443 (1981).
\bibitem{halperin-nelson}
  D.~R.~Nelson and B.~I.~Halperin, Phys.~Rev. B {\bf 19}, 2457 (1979);
  A.~P.~Young, Phys.~Rev. B {\bf 19}, 1855 (1979).
\bibitem{nose}
  S.~Nos\'{e}, J.~Chem.~Phys. {\bf 81}, 511 (1984).
\bibitem{hoover}
  W.~G.~Hoover, Phys.~Rev. A{\bf 31}, 1695 (1985).
\bibitem{cha-fertig}
  M.-C.~Cha and H.~A.~Fertig, Phys.~Rev. B {\bf 50}, 14368 (1994).
\bibitem{allen}
  M.~P.~Allen, D.~Frenkel, W.~Gignac, and J.~P.~McTague, J.~Chem.~Phys.
  {\bf 78}, 4206 (1983).
\bibitem{muto}
  S.~Muto and H.~Aoki, Phys. Rev. B {\bf 59}, 14911 (1999).
\bibitem{footnote1}
  In \cite{schweigert} Schweigert {\it et al.} calculated average
  values for the order parameters, which we think is not sufficient
  for detecting a quasi-long-range order.
\bibitem{footnote2} 
  In \cite{bedanov} Bedanov {\it et al.} found an algebraically decaying
  orientational correlation with an exponent $\approx$0.26 in a region
  $145 < \Gamma < 159$ to conclude the existence of the hexatic phase 
  for a smaller system in microcanonical ensemble, but only thousands
  steps are used for equilibration.
\end{thebibliography}
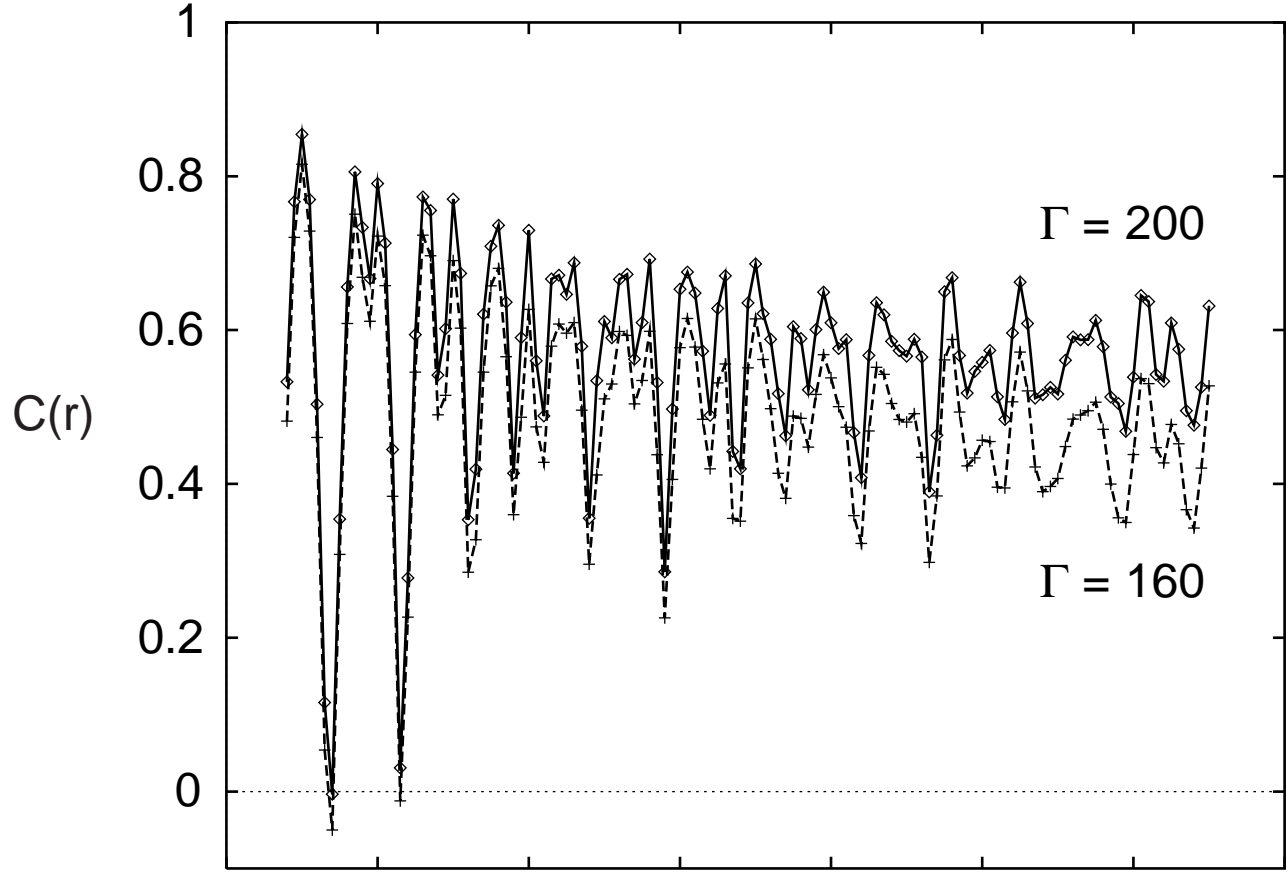
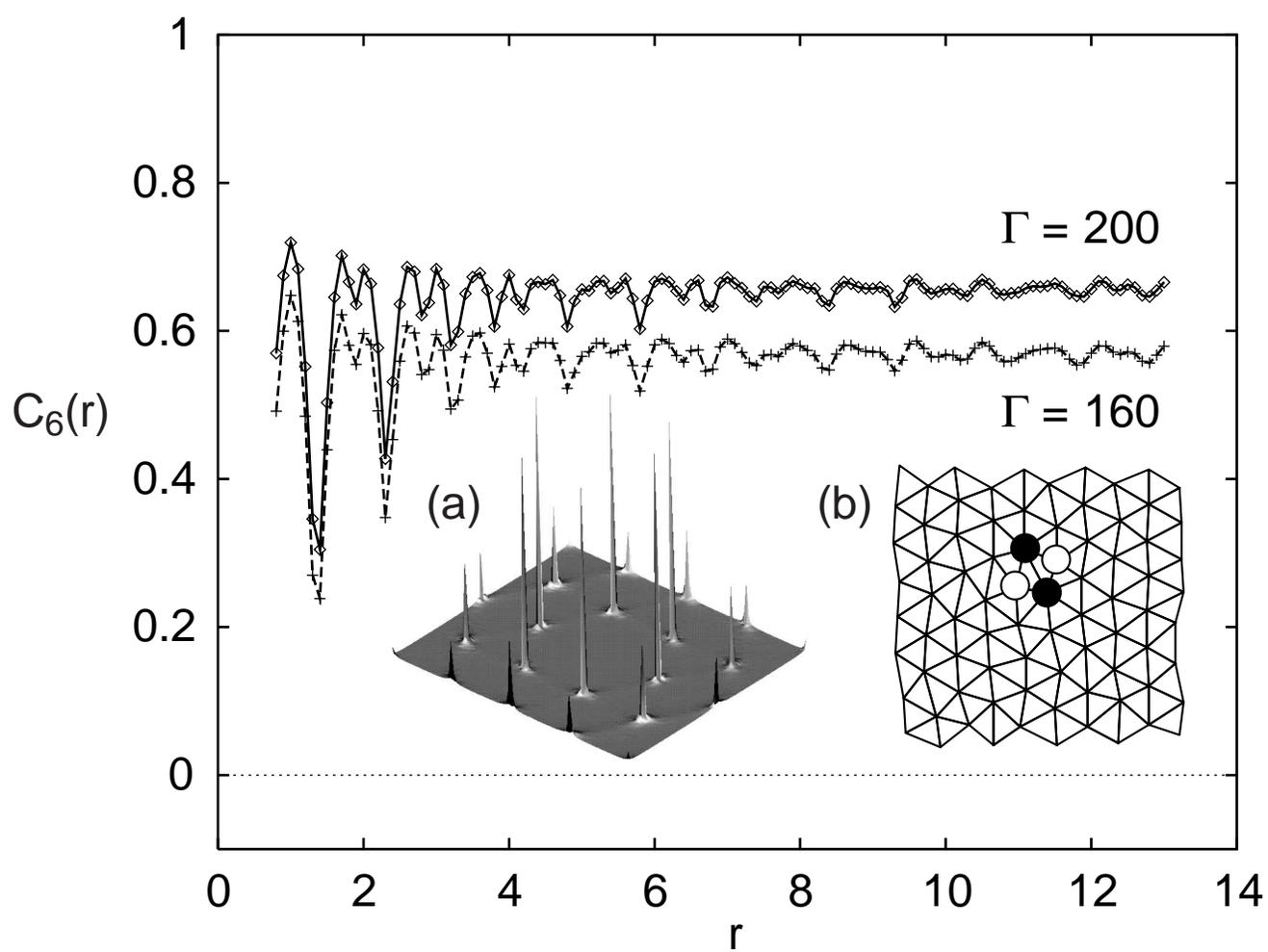
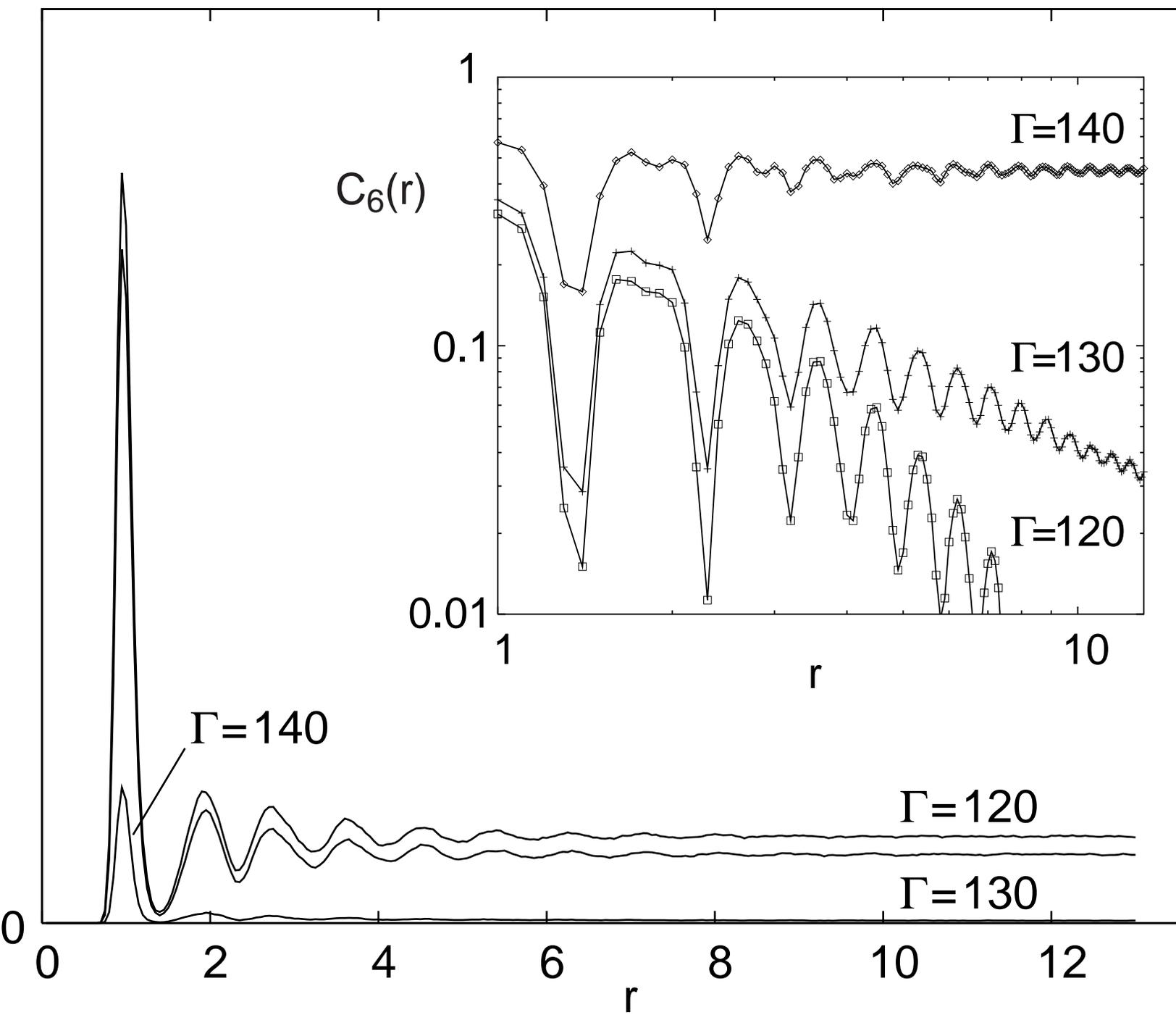
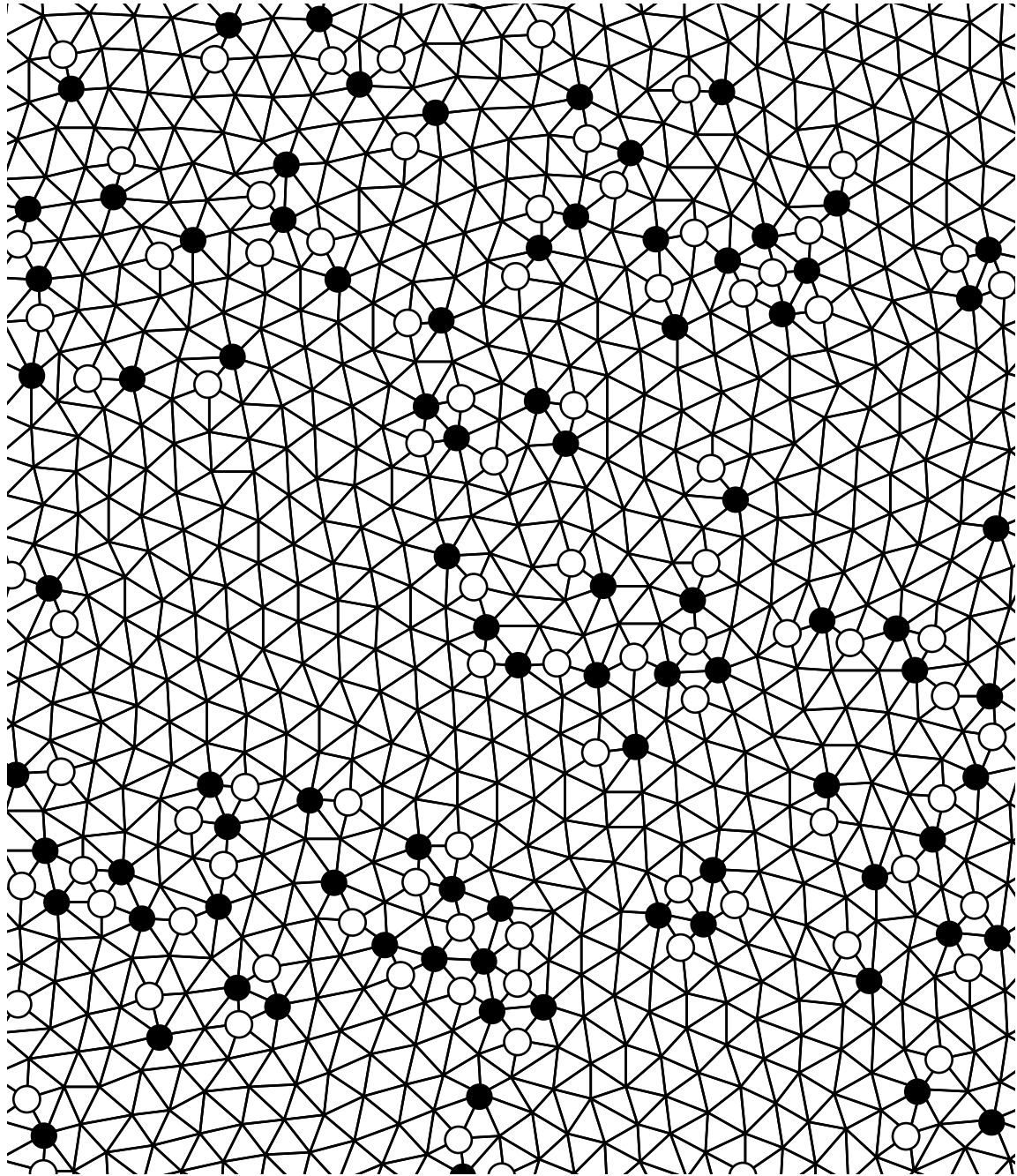
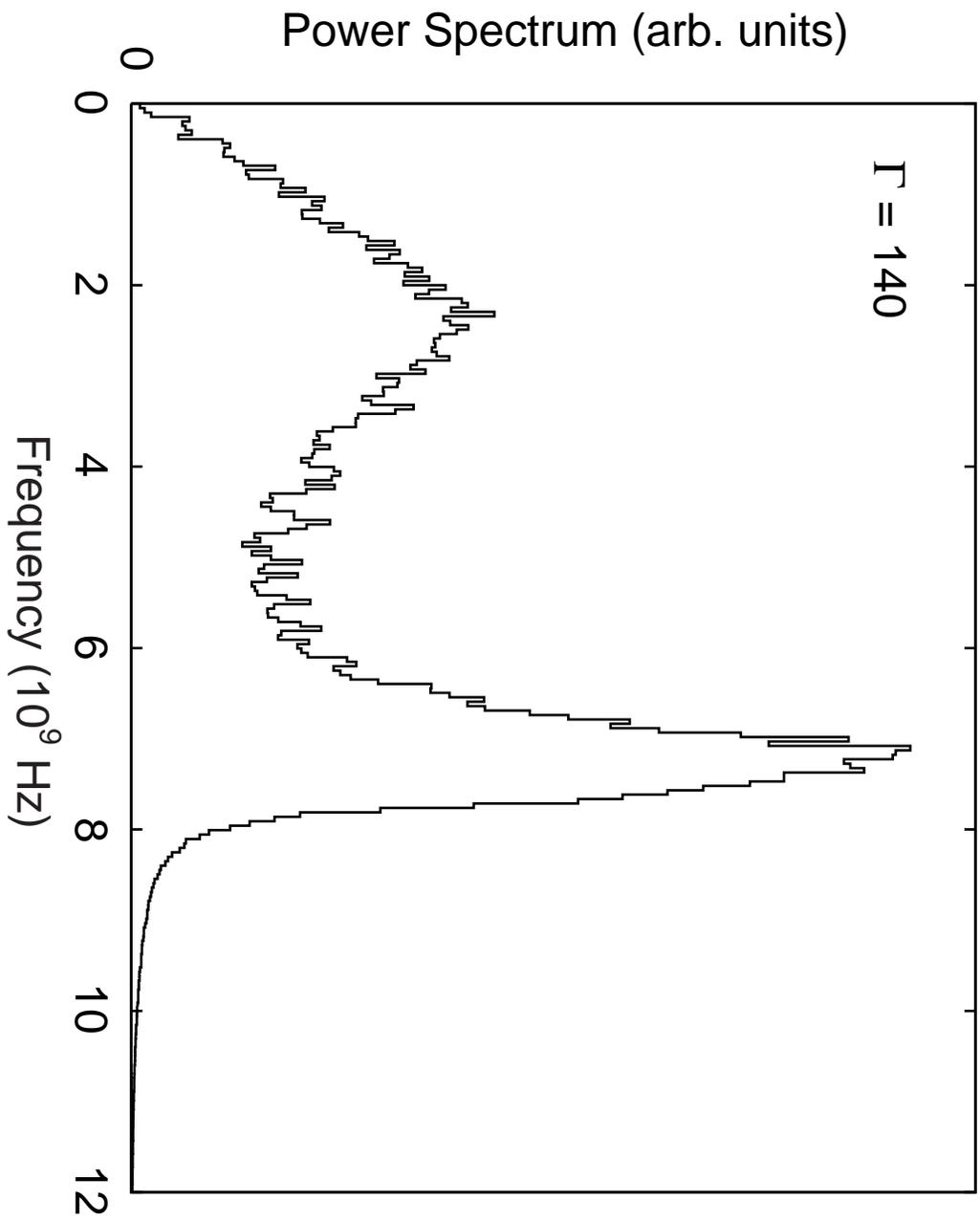
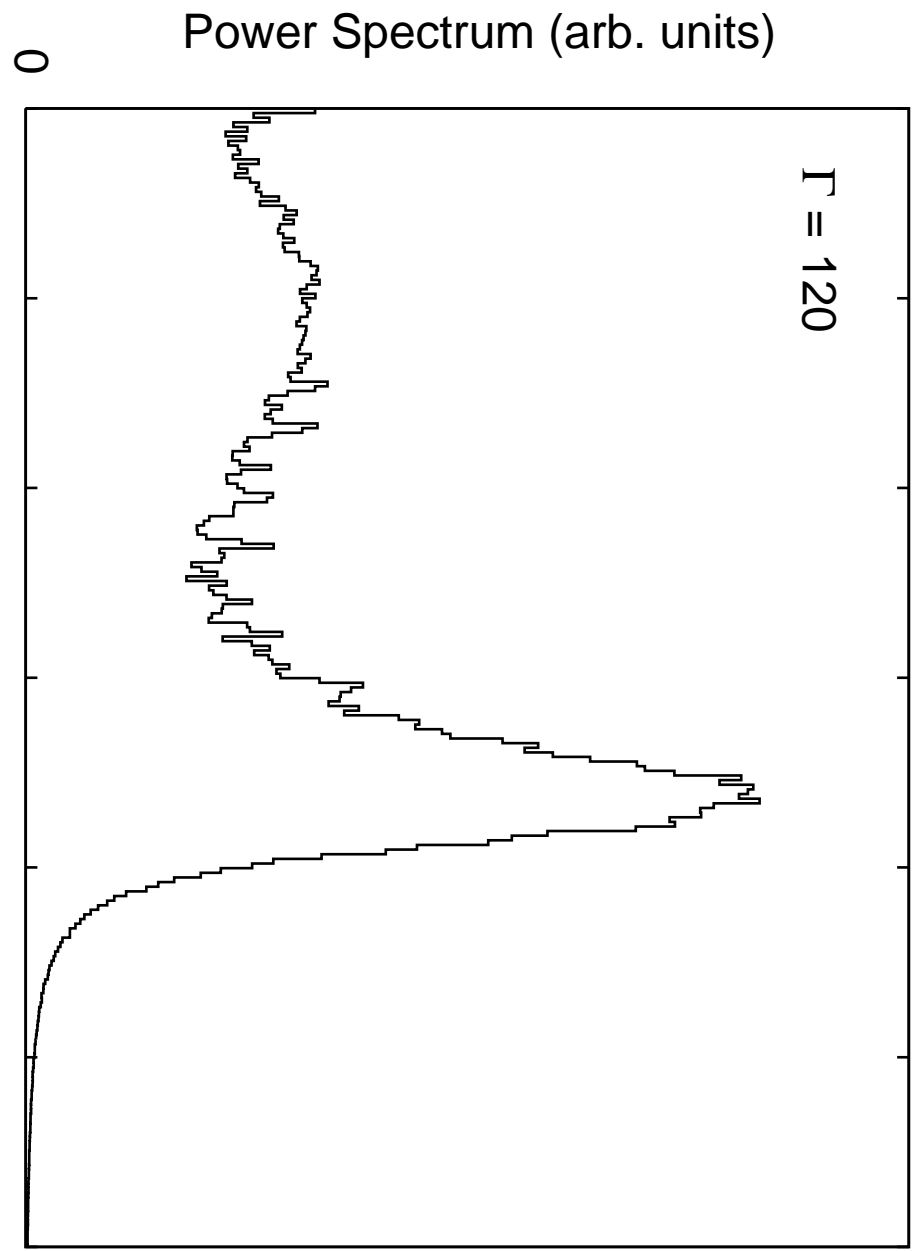
\end{document}